# ON THE MULTI-FACETED SYNERGY BETWEEN X and GAMMA-RAY ASTRONOMIES

Patrizia A. Caraveo
*IASF-INAF, Via Bassini, 15 Milano, ITALY*

When it comes to identifying or to characterizing gamma-ray sources, X-ray observations are of paramount importance. Correlated X-and-gamma-ray flux variations are a powerful identification tool, if the gamma-ray source is unidentified, or an important diagnostic tool to understand the behaviour of an identified source.
Moreover, X-ray observations of non-variable unidentified gamma-ray sources, both galactic and extragalactic, can unveil interesting candidate counterparts, narrowing down the search space and improving significantly the chances for a successful identification.

Swift observations of Fermi gamma-ray-selected pulsar error boxes provide accurate positions of likely counterparts. This makes it possible both to confirm pulsations and to improve timing solution, opening up a new synergy between X and gamma ray observations.

## 1. THE IDENTIFICATION PROBLEM

Historically, the problem of high energy gamma-ray astronomy source identification has been the lack of angular resolution. The resulting positional uncertainties hampered the efforts aimed at identifying gamma-ray sources.

Even with Fermi vastly improved performances[7], the source positional accuracy does not allow for a straightforward identification.

Associations based on positional coincidence have to be confirmed through some supporting evidence.

## 2. VARIABILITY AS AN IDENTIFICATION TOOL

### 2.1. Periodic variabilities: pulsars and binary systems

The presence of a suitable pulsar within a gamma-ray error box is a starting point in the identification process which must rest on the object's time signature.

Folding gamma ray data using the radio timing parameters (usually obtained through ad hoc programmes) is a standard and powerful tool to identify gamma-ray sources. In such a way dozens of radio pulsars have been seen to shine in gamma-rays by Fermi [6] and Agile[13]. The gamma-ray pulsars are young to middle-aged, with rotational energy loss spannig 5 decades, from ~$3 \times 10^{33}$ erg/s to $5 \times 10^{38}$ erg/s, and their apparent efficiencies for conversion to gamma-ray emission range from ~0.1% to unity.

Notably, msec pulsars have been firmly detected by Fermi, establishing them as a new class of near-by gamma-ray emitters [2] characterized, on average, by low rotational energy losses.

Orbital periodic variability has recently been exploited to secure the identification of LSI 61° 303 [5], a potential gamma ray source which has been in the list of suspects since 1981[8], of LS5039 [1] and of Cyg X3[3], also a suspect since the beginning of gamma-ray astronomy.

### 2.2. Random variabilities: blazars

Source confusion and/or association ambiguity is not peculiar to galactic sources in crowded fields.

Also in the extragalactic sky, the positional coincidence of a gamma-ray source with a Blazar only provides a promising association. It becomes an identification if some kind of correlated variability is seen through multiwavelength coverage of the source.

Monitoring such variabilities on different time scales is one of the obvious synergies between gamma-ray astronomy and the rest of the electromagnetic spectrum, exploiting all windows from radio to VHE photons.

Programmed or TOO X-ray monitoring of dozens of blazars is now underway, thanks to the Swift rapid response.

### 2.3. Flaring Galactic sources

Exploiting variability could also provide a way to identify sources seen to briefly shine in our Galaxy by the Agile and Fermi satellites.

Once again TOO observations of variable Fermi and Agile sources are routinely carried out by Swift, providing a much needed coverage in soft and hard X-rays as well as in optical and UV.

Unfortunately, no positive results can be as yet reported. Owing to the limited number of photons available, error boxes for flaring sources are bigger than usual. Such big error boxes deep in the galactic plane make for difficult targets. However, rapid follow-up X-ray observations offer the best (may-be the only) chance to understand such enigmatic sources

## 3. IDENTIFYING STEADY SOURCES

To identify steady gamma-ray sources, X ray coverage is the prime searching tool. Following the chase for Geminga [9], an observational strategy was devised whereby X-ray observations were used to pinpoint potentially interesting sources to be individually characterized and eventually identified through a suite of follow-up X-ray, optical and radio observations.

Requiring several iterations of proposals- approval- data taking- data analysis cycle, such a procedure is cumbersome and long and can be applied only to a limited number of sources. Indeed, a few notable Fermi sources have been (or will be) observed by XMM, Chandra and Suzaku.

The use of powerful X-ray observatories is mandatory in order to collect enough statistics to characterize the newly found sources through their spectra and to make it possible to search for periodic, pulsar-like, variability. Once again, such an approach follows the Geminga paradigm, where X-ray observations were instrumental in finding the source pulsations, later to be confirmed also in gamma-rays, paving the way to the source identification.

Although applied to a number of EGRET sources, such Geminga paradigm did not result in new identifications, see e.g. [12]. While a number of promising, pulsar-like X-ray candidates where unveiled, none yielded the time signature needed to confirm their isolated neutron star nature. Indeed, their X-ray counterparts were quite faint, rendering very difficult the search for pulsation.

At the time of the Fermi launch, efforts to identify gamma-ray galactic sources were at a dead end.

## 4. TOWARD A NEW SYNERGY

### 4.1. Position, position, position

After only a few weeks of operations, Fermi changed our perspective by discovering the pulsation of the source associated with CTA-1, at the position of its proposed X-ray counterpart [11]. Showing for the first time that the gamma timing signature can be found directly from the gamma ray data, Fermi provided a big step forward in the understanding of galactic unidentified sources[4]. However, the limited accuracy of gamma-ray positions hampers Fermi periodicity searches. While the timing signature of a source can be easily found using a week time span, it is difficult to keep track of the pulsar phase over a much longer time. To preserve the phase over a time span of months, photon arrival times must be precisely barycentrized, a standard procedure whose accuracy depends on the knowledge of the position of the spacecraft as well as on the source positioning. A rough source position translates into a rough barycentric correction which could weaken or even destroy the phase coherence of the source photons. For gamma-ray pulsars found inside EGRET sources already extensively mapped, the position of the proposed counterparts were used, capitalizing on the multiwavelength work done in the past decade by a number of authors.

For new sources, the most likely position could be searched by trying several choices inside the gamma-ray error box. However, apart from being time consuming, such a procedure multiplies the number of trials, thus decreasing the overall pulsation significance.

Pinpointing likely X-ray counterparts of Fermi gamma-ray selected pulsars, i.e. isolated neutron stars (INSs) not detected at radio wavelengths, has proven very helpful [4]. Swift observations of Fermi INS error boxes were quickly performed, unveiling one or more counterparts for half of the gamma INSs. When the SWIFT coverage yielded more than one candidate, the timing analysis was used to select the best one, or to discard all candidates if none yielded a significant improvement to the overall timing solution.

## 5. CONTINUING THE OLD FASHIONED WAY

While such a procedure opens up a new, powerful synergy between X and gamma ray observations, based solely on source positions, the full characterization of the X-ray behaviour of a radio quiet, gamma-ray-selected pulsar must rely on the data harvest of instruments such as XMM-Newton, Chandra or Suzaku.

A 130 ksec long XMM-Newton observation of CTA-1 provides a very good example of the challenges we are facing[10].

The gamma-ray source X-ray counterpart is detected together with its PWN and the surrounding SNR. While the source spectral analysis clearly shows the presence of a thermal component superimposed to a non-thermal one, typical of an INS emission, the standard periodicity search does not yield any significant result.

To detect the X-ray pulsation one has to capitalize on Fermi timing solution which yields a highly significant light curve in soft X-rays.

The case of CTA-1 makes it clear that, when searching for periodicity of unidentified source, X-rays are not the driving tool any more.

### Acknowledgments

This work was made possible by the contributions of Andrea De Luca, Martino Marelli, Andrea Belfiore (IASF-INAF, Milano), Pablo Saz Parkinson (UCSC) and Paul Ray (NRL).

We are indebted to Neil Gehrels for his continuous support in using SWIFT in conjunction to Fermi and Agile.

This work is supported by ASI-INAF contracts I/047/08/0 and I/011/07/0

**Insert PSN Here**

2-